\documentclass[prb,twocolumn,aps]{revtex4-2}
\usepackage{graphicx}
\usepackage{dcolumn}
\usepackage{bm}
\usepackage{amssymb}
\usepackage{amsmath}
\usepackage{physics}
\usepackage{sublabel}
\usepackage{subfigure}
\usepackage[utf8]{inputenc}
\usepackage{hyperref}
\usepackage[usenames, dvipsnames]{color}
\usepackage[english]{babel}
\usepackage[T1]{fontenc}
\usepackage{bbm}

\usepackage{float}
\usepackage{verbatim}
\usepackage[most]{adjustbox}
\hypersetup{colorlinks = true, linkcolor=blue, citecolor=blue, urlcolor=blue}

  \makeatletter
    \renewcommand\@make@capt@title[2]{%
     \@ifx@empty\float@link{\@firstofone}{\expandafter\href\expandafter{\float@link}}%
      {\textsc{#1}}\@caption@fignum@sep#2\quad}%
    \makeatother

\begin{document}
\title{Giant excitonic magneto-optical Faraday rotation in single semimagnetic $\mathbf{CdTe/Cd_{1-x}Mn_{x}Te}$ quantum ring}
\author{Kalpana Panneerselvam}
\author{Bhaskaran Muralidharan}
\thanks{corresponding author:bm@ee.iitb.ac.in}
\affiliation{Department of Electrical Engineering, Indian Institute of Technology Bombay, Powai, Mumbai-400076, India}

\date{\today}
\begin{abstract}
Magnetic tuning of the bound exciton states and corresponding giant Zeeman splitting (GZS) between $\mathrm{\sigma^{+}}$ and $\mathrm{\sigma^{-}}$ excitonic transitions in $\mathrm{CdTe/Cd_{1-x}Mn_{x}Te}$ quantum ring has been investigated in the Faraday configuration for various concentrations of $\mathrm{Mn^{2+}}$ ions, using the variational technique in the effective mass approximation. The sp-d exchange interaction between the localized magnetic impurity ions and the delocalized charge carriers has been accounted via mean-field theory with the inclusion of a modified Brillouin function. The enhancement of the GZS, and in turn, the effective g-factor with the application of an external magnetic field, is strikingly manifested in type-I – type-II transition in the band structure, which has been well explained by computing the overlap integral between the electron and hole, and the in-plane exciton radius. This highlights the extraordinary magneto-optical properties, including the giant Faraday rotation and associated Verdet constant, which have been calculated using single oscillator model. The oscillator strength and exciton lifetime have been estimated, and are found to be larger than in the bulk diluted magnetic semiconductors (DMS) and quantum wells, reflecting stronger confinement inside the quantum ring. The results show that the DMS-based quantum ring exhibits more extensive Zeeman splitting, which gives rise to ultra-high Verdet constant of $\mathrm{2.6 \times 10^{9}rad/Tesla/m}$, which are a few orders of magnitude larger than in the existing quantum systems and magneto-optical materials. \end{abstract}


\maketitle

\section{Introduction}\label{1}
Diluted magnetic semiconductors (DMS) are of special interest because of their unique combination of semiconducting and magnetic properties. The sp-d exchange interaction between the localized magnetic moments of the dopant magnetic ions and the spins of the charge carriers (electrons/holes) in DMS \cite{gaj1993relation, ivanov2008optically, rice2017direct, kalpana2017donor, anitha2020magnetization, kozyrev2021optical} significantly alters the energy spectra of the carriers, which greatly enhances the spin-dependent effects. Moreover, these effects can be widely tuned by an external magnetic field, temperature, and the concentration of magnetic ions to induce fascinating magneto-optical (MO) and magneto-electrical properties. Among all the exciting signatures of such exchange interaction, the striking consequences are the giant Zeeman splitting (GZS) \cite{kuhn1994zeeman, fainblat2016giant, barrows2017excitonic} and the giant Faraday rotation (GFR) \cite{gaj1993giant,panmand2020characterisation}. The Zeeman splitting between the band states with different spin components generates the spin – polarization of the conducting carriers, which is exploited as spin aligners in spintronic devices \cite{schmidt2001dilute, ferrand2001applications}, an anomalous magnetoresistance at low temperature \cite{fukumura1999oxide}, and vastly amplifies the conversion of the spin current into an electrical current \cite{ganichev2009spin}. Hence, the DMS have been an active area of research as an alternative to the ferromagnetic metal contacts for the efficient spin-injection into non-magnetic semiconductors, spin detection, and the realization of the spin-polarized transport in semiconductor structures, which have substantial industrial applications in the field of magnetoelectronics, spintronics, and solid-state quantum computing \cite{hirase2019carrier, yahia2010p, kanaki2018large, terada2022bias, ohno1999electrical, moro2014spin, kobak2014designing}. The concept of the FR (a solid rotation of the plane of polarization of light travels in a magnetized medium along the applied magnetic field) manifests itself in various MO devices such as optical isolators, Faraday rotators, and optical circulators for high-speed optical communication systems \cite{turner1983new, ju2019temperature, panmand2020characterisation, carothers2022high}, for which DMS act as potential MO materials.\\
\indent The carrier localization and its transport properties have been examined using DMS materials in various nanostructured systems like quantum wells, wires, and dots \cite{oka2002magneto, chang2003spin, awschalom1999spin, chang2003magnetic}. Considerable attention paid to the quantum ring (QR)-based infrared photodetectors and lasers \cite{le2018colloidal, samadzadeh2015tunable} in recent times due to its doubly-connected topological nature has engendered an interest in us to study how the radial and axial confinement of the individual carriers and the exciton in semimagnetic QR  impact the sp-d exchange interaction in the Faraday configuration (the magnetic field is applied along the direction of observation (z) and parallel to the light wave vector). The strong sp-d coupling makes magnetic ions mediate the influence of the magnetic field on the band gap engineering by enhancing the Zeeman splitting of the energy levels, which is strikingly manifested in type-I - type-II transition in the band offset \cite{klar1997magnetic, deleporte1990magnetic, delalande1992excitons, kuroda1996magnetic}. Hence, the DMS extends its potential applications to optoelectronics due to the possible tuning of the band states, which in turn tune the emission wavelength widely over Near- to Far-IR, creating a giant optical response.\\
\indent This article aims to delineate the magnetic tuning of exciton energy states due to the GZS between $\mathrm{\sigma^{+}}$ and $\mathrm{\sigma^{-}}$ spin components, in turn, the effective g-factor for various mole fractions of (x) magnetic dopants in $\mathrm{CdTe/Cd_{1-x}Mn_{x}Te}$ QR since CdMnTe has well served as a potential MO material for the past few decades towards optoelectronic applications. Various MO parameters, such as oscillator strength, radiative lifetime, and radiative decay rate, have been evaluated. The occurrence of type-I - type-II transition in a single semimagnetic QR has been well explained in the present communication by computing the overlap integral between the electron and hole, and estimating the in-plane exciton radius. Though QRs are more flexible for experimental developments \cite{kleemans2007oscillatory, bayer2003optical, ding2010gate, hackens2006imaging, le2018colloidal} due to the advances in fabrication procedures, and DMS addresses the fundamental challenges in the spintronic devices in its unique way, the possible integration of DMS into QR structures has not yet been developed to unveil the hidden mystery. Few theoretical studies have been proposed on single and concentric double QRs doped with transition metal ions focusing on the magnetic and thermal properties \cite{kalpana2020impurity, babanli2020aharonov, janet2021diluted} but not on the MO properties of excitons, which is of novel interest in the present work.\\ 
\indent We later show a theoretical evaluation of the Verdet constant of a remarkable MO phenomenon, the GFR, using single oscillator model. The source for the larger values of the Verdet constant in DMS has been traced down to the GZS of the energy band states near the band gap resonance. Although most research has focused on achieving a larger Verdet constant with various MO materials, especially $\mathrm{Cd_{1-x}Mn_{x}Te}$, these studies have been restricted only to bulk DMS \cite{gaj1993giant, bartholomew1986interband, jimenez1992near, hwang2006temperature, hugonnard1994faraday} and epitaxial heterostructures in the form of wells \cite{buss1997giant, nakamura1990faraday, kohl1991faraday, buss1995excitonic, gourdon2002enhanced} and dots \cite{nelson2015picosecond, panmand2020characterisation, carothers2022high}. Therefore, investigating the impact of sp-d exchange interaction on the GFR in DMS heterostructures with various topologies, like QR, would generate unprecedented interest in developing high-quality epitaxial structures for various technological applications. \\
\indent In the following, sec. \ref{2} discusses the theoretical formalism using the variational technique in the effective mass approximation to solve for bound exciton states in $\mathrm{CdTe/Cd_{1-x}Mn_{x}Te}$ QR at liquid helium temperature. The mean-field theory with the modified Brillouin function to account for sp-d exchange interaction is explained in detail, and the occurrence of the GZS in nanostructures is also delineated. The results of variation of interband transition energy, the binding energy of $\mathrm{\sigma^{\pm}}$ magneto-exciton, and various MO properties, including GZS and GFR in QR doped with various concentrations of $\mathrm{Mn^{2+}}$ ions (x = 0.5\%, 1\%, 5\%, 10\%, and 20\% of $\mathrm{Mn^{2+}}$ ions) are presented and discussed in sec. \ref{3.1}-\ref{3.5}. Section \ref{4} elucidates the significance of the experimental validation in QR based on DMS by comparing the present results with those already reported for the bulk and QWs.

\section{Theoretical Model }\label{2}

\begin{figure*}[!htbp]
   \centering
   \includegraphics[width=12cm, height=10cm]{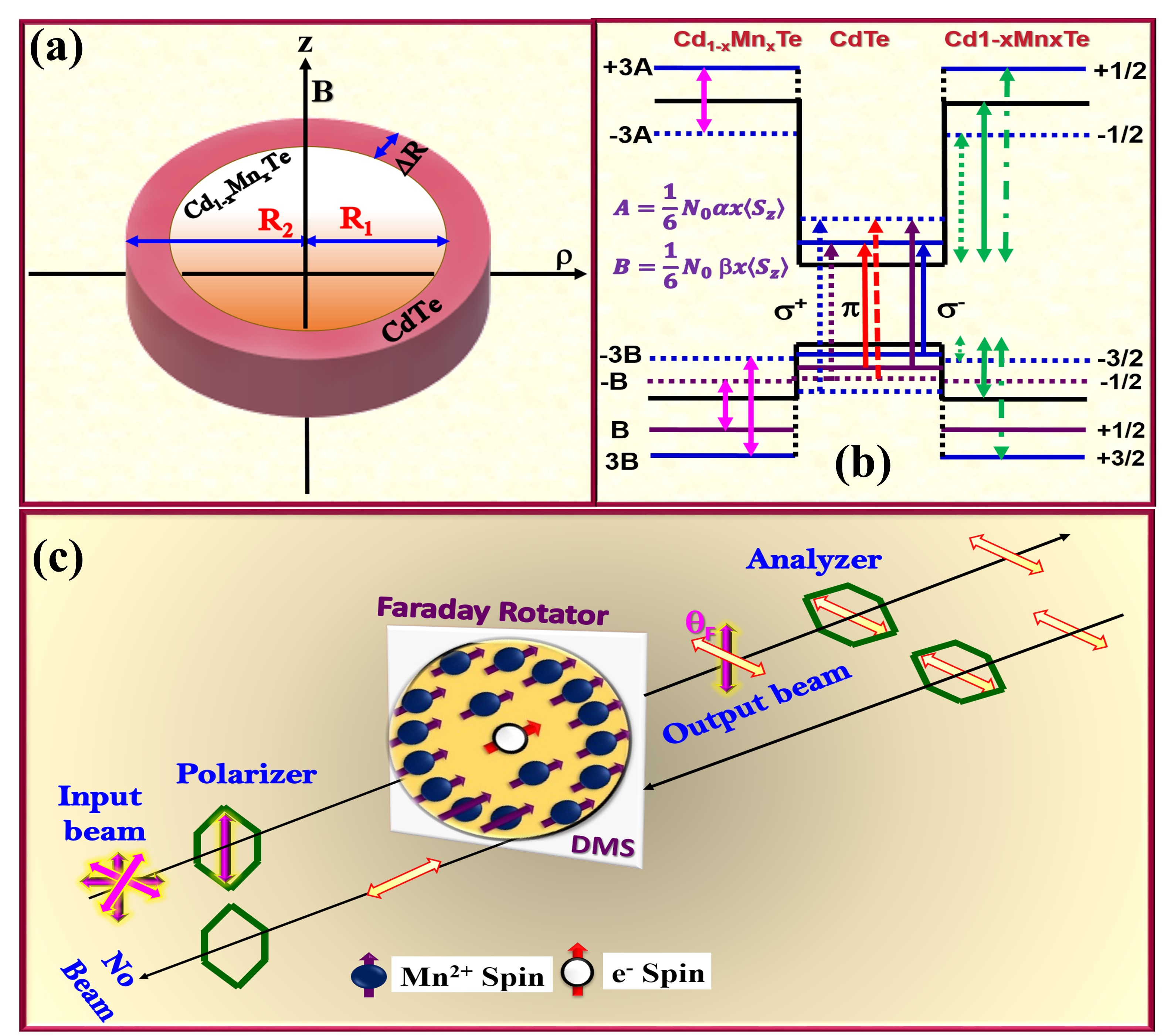} 
    \caption{ Schematics: (a) Profile of the $\mathrm{CdTe/Cd_{1-x}Mn_{x}Te}$ SQR. (b) Giant Zeeman splitting of excitonic energy levels in $\mathrm{CdTe/Cd_{1-x}Mn_{x}Te}$ and corresponding optical transitions ($\mathrm{\sigma^{+}, \sigma^{-}, \pi}$). (c) The concept of Faraday rotation in DMS SQR.}
    \label{Fig:1}
 \end{figure*}
 The schematic diagram of a single quantum ring (SQR) is displayed in (Fig. \ref{Fig:1}(a)). The Schrodinger equation and corresponding Hamiltonian for the ground state bound electron-hole pair subjected to a magnetic flux in DMS SQR is written in a dimensionless form, considering the effective Rydberg ($\mathrm{R^{*}}$) as a unit of energy and effective Bohr radius ($\mathrm{a_{B}^{*}}$) as a unit of length, and is given by,

\begin{subequations}
\begin{equation}
\begin{aligned}
&\hat{H}_{ex} \Psi_{ex} = E_{ex} \Psi_{ex}\\ 
\end{aligned}
\label{eqn:1a}
\end{equation}
\begin{equation}
\begin{aligned}
 &\hat{H}_{ex} = -\frac{1}{\rho_{e}^2} \frac{\partial^{2}}{\partial \varphi^{2}} -\frac{1}{\rho_{h}^2} \frac{\partial^{2}}{\partial \varphi^{2}} -\frac{\mu(T)}{m_{e}^{*}(T)} \left(\nabla \rho_{e}^{2} + \nabla z_{e}^{2} \right)\\
   &\quad -\frac{\mu(T)}{m_{h}^{*}(T)} \left(\nabla \rho_{h}^{2} + \nabla z_{h}^{2} \right) + V_{B} (\rho_{e},z_{e}) + V_{B} (\rho_{h},z_{h})\\
 &-\frac{e^{2}}{\epsilon (T) |\Vec{r_{e}}-\Vec{r_{h}}|} +i \, \gamma \frac{m_{h}^{*}-m_{e}^{*}}{m_{h}^{*}+m_{e}^{*}} \frac{\partial}{\partial \varphi} + \frac{\gamma^{2} \rho^{2}}{4}
\end{aligned}
\label{eqn:1b}
\end{equation}
\end{subequations}

\noindent where, e and h represent the electron and hole, respectively. The strength of the magnetic field is parametrized by $\mathrm{\gamma = \frac{\hbar \omega_{c}}{2 R^{*}}}$, $\mathrm{\omega_{c}}$ is the cyclotron frequency. Since the electron and hole move freely along the annular part of the ring, their motions no longer depend on $\mathrm{\phi_e}$ and $\mathrm{\phi_h}$ separately, but on the relative angular displacement $\mathrm{\phi = \phi_e-\phi_h}$ and it should be treated with the reduced effective mass ‘$\mu$’ of the exciton. Moreover, the material parameters, effective mass, and spatial dielectric constant are considered as temperature-dependent.The sp-d exchange interaction between the electron (hole) and the localized $\mathrm{Mn^{2+}}$ magnetic dopants is denoted by $\mathrm{\hat{H}_{sp-d}}$, and is written as \cite{gaj1993relation, kalpana2017donor, kalpana2019magnetic},

\begin{equation}
\begin{aligned}
\hat{H}_{sp-d} &= -\sum_{i} J(\boldsymbol{r_{e}} - \boldsymbol{R_{i}} ) \boldsymbol{\hat{S}_{i}} \cdot \boldsymbol{\hat{s}_{e}}-\sum_{i} J(\boldsymbol{r_{h}} - \boldsymbol{R_{i}} ) \boldsymbol{\hat{S}_{i}} \cdot \boldsymbol{\hat{s}_{h}}
\end{aligned} 
\label{eqn:2}
\end{equation}

\noindent ‘J’ is the coupling constant for the exchange interaction between the electron (hole) of spin $\mathrm{\hat{s}_{e}}$ ($\mathrm{\hat{s}_{h}}$) located at $\mathrm{\boldsymbol{r_{e}}}$ ($\mathrm{\boldsymbol{r_{h}}}$) and the spin $\mathrm{\hat{S}_{i}}$ of the $\mathrm{Mn^{2+}}$ ions located at sites $\mathrm{\boldsymbol{R_{i}}}$. $\mathrm{V_{B}(\rho_{e,h},z_{e,h})}$ in Eq. (\ref{eqn:1b}) is the confining potential of the SQR and is modeled by an abrupt square potential:

\begin{equation}
\begin{aligned}
    V_{B}(\rho_{e,h},z_{e,h}) = \begin{cases} 0 &  R_{1} < \rho_{e,h} \leq R_{2},\\
                                                & \ -d/2 < z_{e,h} \leq +d/2 \\
    V_{0 e,h} & \mathrm{otherwise}
    \end{cases}
    \end{aligned}
    \label{eqn:3}
\end{equation}

 \noindent $\mathrm{V_{0 e} = 70\% \Delta E_{g}^{B}}$ , and $\mathrm{V_{0 h} = 30\% \Delta E_{g}^{B}}$ represent the potential band offset formed in the conduction and valence bands, respectively. Tuning of the potential barrier height, $\mathrm{V_{0 e}}$ and $\mathrm{V_{0 h}}$ with the applied field, $\mathrm{B_{z}}$, is possible due to the Zeeman splitting of the band edges (Fig. \ref{Fig:1}(b)) and is written by a formula suggested by K. Navaneethakrishnan et al \cite{jayam2002optical} that satisfactorily fits the experimental Zeeman splitting values available for the $\mathrm{Mn^{2+}}$ compositions x = 0.07, 0.24, and 0.3 with a maximum error of 5\%. Hence, the same formula is adopted here, and the fitting equation is given by \cite{jayam2002optical,gnanasekar2004spin,kalpana2017donor, kalpana2019magnetic},
 
\begin{equation}
\begin{aligned}
\Delta E_{g}^{B} &= \Delta E_{g}^{0} \, \frac{\eta_{e,h} \, e^{\zeta_{e,h} \, \gamma} - \, 1}{\eta_{e,h} - 1}
\end{aligned}
\label{eqn:4}
\end{equation}

\noindent $\mathrm{\Delta E_{g}^{B}}$ and $\mathrm{\Delta E_{g}^{0}}$ denotes the band gap difference between the well CdTe layer and the barrier $\mathrm{Cd_{1-x}Mn_{x}Te}$ layer in the presence and absence of applied magnetic field, respectively. $\mathrm{\eta_{e,h} = e^{\zeta_{e,h} \, \gamma_{0}}}$ is chosen with a fitting parameter $\mathrm{\zeta_{e} (\zeta_{h}) = 0.5 (-0.5)}$, and $\mathrm{\gamma_{0}}$ is a critical magnetic field at which the barrier completely vanishes. The critical magnetic field $\mathrm{\gamma_{0}}$ in Tesla for different magnetic dopant compositions is given for conduction (valence band) as $\mathrm{\gamma_{0}} = A \, e^{nx}$ with A = 0.734 and n = 19.082 (A = - 0.57 and n = 16.706).The most appropriate trial wavefunction of a ground state exciton is written in a non-separable form due to correlated electron-hole pair as,

\begin{equation}
\begin{aligned}
\Psi_{ex}(r_{e}, r_{h}) &= N_{1s} \, \phi_{e} (\rho_{e}) \, \phi_{h} (\rho_{h}) \, f_{e}(z_{e}) \, f_{h}(z_{h}) e^{-\lambda \, \boldsymbol{r_{eh}}}
\end{aligned}
\label{eqn:5}
\end{equation}

\noindent where, $\mathrm{\phi_{e,h} (\rho_{e,h})}$ and $\mathrm{f_{e,h} (z_{e,h})}$ are the envelope functions along the radial and axial directions, respectively. $\mathrm{e^{-\lambda \, \boldsymbol{r_{eh}}}}$ describes the correlation between the electron and hole which depends mainly on the distance between the two. $\mathrm{r_{eh} = \sqrt{|(\rho_{e}-\rho_{h})|^{2}+|(z_{e}-z_{h})|^{2}}}$, whereas, $\mathrm{|(\rho_{e}-\rho_{h})|^{2}}$ denotes the projection of the distance between the electron and hole on the plane of the QR and is given by, $\mathrm{|(\rho_{e}-\rho_{h})|^{2}} = (\rho_{e}^{2} + \rho_{h}^{2} -2 \rho_{e} \rho_{h} \cos(\varphi))^{1/2}$.

\begin{subequations}
    \begin{equation}
\begin{aligned}
    \phi (\rho_{e,h}, \varphi_{e,h}) = \begin{cases} \phi _{I} (\rho_{e,h}),  & \rho_{e,h} \leq R_{1} \\
    \phi _{II} (\rho_{e,h}), & R_{1} < \rho_{e,h} \leq R_{2}\\
    \phi _{III} (\rho_{e,h}), & \rho_{e,h} > R_{2}\\
   \end{cases}
    \end{aligned}
    \label{eqn:6a}
\end{equation}
\begin{equation}
\begin{aligned}
    & \phi _{I} (\rho_{e,h}) =  C_{1,e,h} \, I_{0} \left( \beta_{e,h}, \rho_{e,h} \right)   \\
    & \phi _{II} (\rho_{e,h}) = C_{2,e,h} \, J_{0} \left( \alpha_{e,h}, \rho_{e,h} \right) + C_{3,e,h} \, Y_{0} \left( \alpha_{e,h}, \rho_{e,h} \right) \\
    & \phi _{III} (\rho_{e,h}) = C_{4,e,h} \, K_{0} \left( \beta_{e,h}, \rho_{e,h} \right) 
    \end{aligned}
    \label{eqn:6b}
\end{equation}
\begin{equation}
\begin{aligned}
    f (z_{e,h}) = \begin{cases} B_{e,h} \, \exp[k_{e,h} \, z_{e,h}],  &  -\infty < z_{e,h} \leq -d/2 \\
    \cos(\kappa_{e,h} \, z_{e,h}), & -d/2 < z_{e,h} \leq +d/2 \\
    B_{e,h} \, \exp[-k_{e,h} \, z_{e,h}],  &  +d/2 < z_{e,h} \leq +\infty
    \end{cases}
    \end{aligned}
    \label{eqn:6c}
    \end{equation}
\end{subequations}

\noindent where, $\mathrm{\beta_{e,h} = \frac{m_{b}^{*} (V_{0 e,h}- E_{\rho_{e,h}})}{\hbar^{2}} \; ; \; \alpha_{e,h} = \frac{m_{w}^{*} E_{\rho_{e,h}}}{\hbar^{2}}}$ \\

$\mathrm{k_{e,h} = \frac{m_{b}^{*} (V_{0 e,h}- E_{z_{e,h}})}{\hbar^{2}} \; ; \; \kappa_{e,h} = \frac{m_{w}^{*} E_{z_{e,h}}}{\hbar^{2}}}$ \\ 

\noindent Here, $\mathrm{N_{1s}}$ is the normalization constant. $\mathrm{C_{1,e,h}}$, $\mathrm{C_{2,e,h}}$, $\mathrm{C_{3,e,h}}$, and $\mathrm{C_{4,e,h}}$ are obtained by choosing proper boundary conditions. $\mathrm{E_{\rho e,h}}$, $\mathrm{E_{z e,h}}$ are the subband energy levels formed due to the radial and axial confinement of the QR. Invoking the variational technique, the binding energy ($\mathrm{E_{B_{ex}}}$) and the interband transition energy ($\mathrm{E_{T_{ex}}}$) of the exciton is computed using the form,

\begin{equation}
\begin{aligned}
    &E_{B_{ex}} = E_{\rho_{e}} + E_{\rho_{h}} + E_{z_{e}} + E_{z_{h}} + \gamma - \langle \boldsymbol{H_{ex}} \rangle _{min}\\
    &E_{T_{ex}} = E_{g}(T) + \langle \boldsymbol{H_{ex}} \rangle _{min}
\end{aligned}
\label{eqn:7}
\end{equation}

\begin{figure*}[!htbp]
   \centering
    \includegraphics[width=1.01\linewidth]{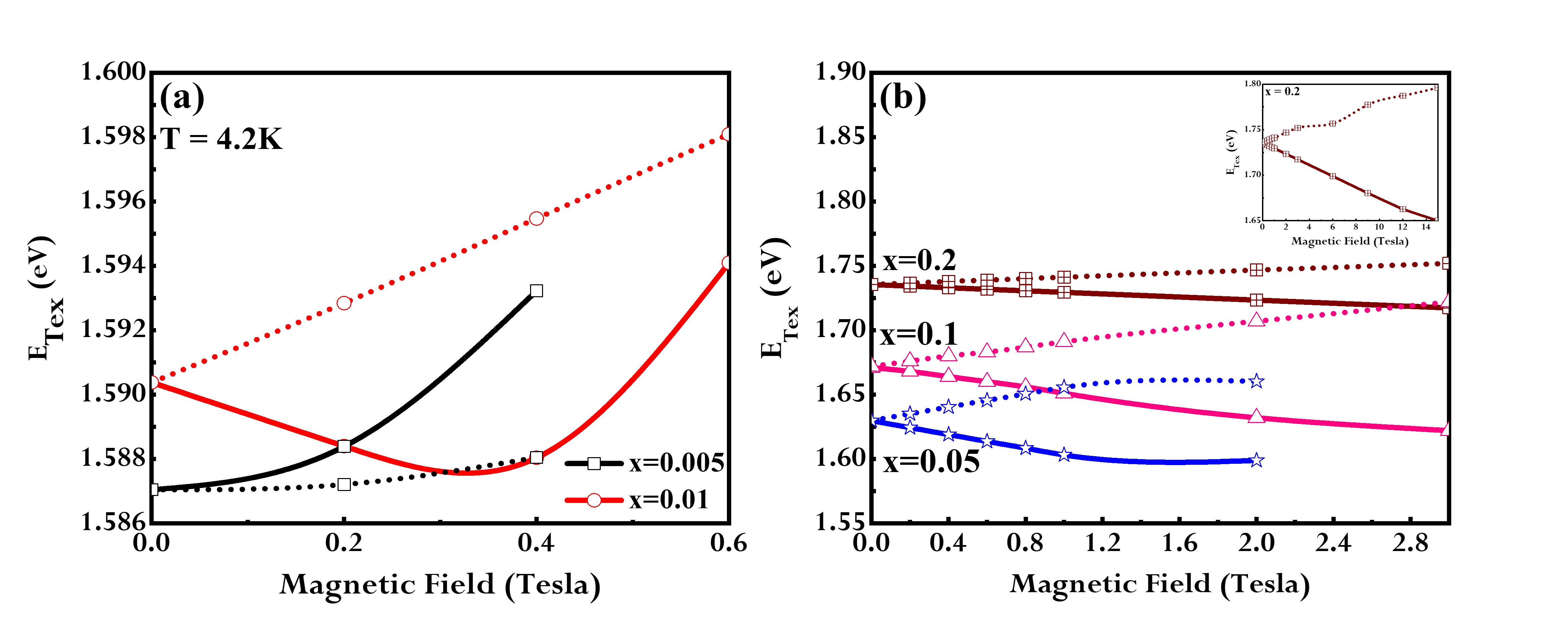}
    \caption{Interband transition energy as a function of magnetic field for $\mathrm{\sigma^{+}}$ and $\mathrm{\sigma^{-}}$ exciton for various dopant concentrations. (a) $\mathrm{x \leq 0.01}$ and (b) x > 0.01.}
    \label{Fig:2}
   \end{figure*}
   
\noindent In the Faraday geometry, the magnetic moments of the ensemble of $\mathrm{Mn^{2+}}$ ions with spin angular momentum $\mathrm{S_{Mn} = 5/2}$ are subjected to the sp-d exchange interaction with the conduction band electrons of spin s = 1/2 and the heavy hole valence band with angular momentum J = 3/2. This causes the heavy hole exciton splits into two components with angular momentum +1 and -1, which is composed of $\mathrm{s_{z} = -1/2}$, $\mathrm{J_{z} = +3/2}$ and $\mathrm{s_{z} = 1/2}$, $\mathrm{J_{z} = -3/2}$, respectively. The GZS between the excitonic energy levels exhibited in the nanostructures is as similar as in bulk DMS, but with a difference in the potential barrier height experienced by the two different spin states. The schematic diagrams which explain the Zeeman splitting of the energy levels in DMS nanostructures and its resultant GFR are depicted in Fig. \ref{Fig:1}(b) and Fig. \ref{Fig:1}(c). The applied magnetic field increases and decreases the potential barrier for the spin up and spin down states, respectively, and thereby the corresponding confinement of both the electron and heavy hole with $\mathrm{s_{z} = +1/2}$, $\mathrm{J_{z} = +3/2}$, and $\mathrm{s_{z} = -1/2}$, $\mathrm{J_{z} = -3/2}$ becomes stronger and weaker, respectively, inside the QR. Therefore, by magnetically tuning the potential barrier, the energy levels of the exciton inside the QR could also be tuned, manifesting itself in two different excitonic transitions, namely $\mathrm{\sigma^{+}}$ and $\mathrm{\sigma^{-}}$. $\mathrm{\sigma^{+}}$ corresponds to the transition between $\mathrm{J_{z} = -3/2}$ (heavy hole) and $\mathrm{s_{z} = -1/2}$ (electron) states, and $\mathrm{\sigma^{-}}$ transition involves $\mathrm{J_{z} = +3/2}$ and $\mathrm{s_{z} = +1/2}$ states. The splitting of the energy level corresponding to two transitions is expressed by 
\cite{gaj1993relation},

\begin{equation}
\begin{aligned}
    E_{\pm} &= \pm \frac{1}{2} x_{eff} N_{0} \left(\beta_{exc}-\alpha_{exc}\right) \left \langle S_{z}^{Mn}(B)\right \rangle
\end{aligned}
\label{eqn:8}
\end{equation}

\noindent The Zeeman splitting energy between the two excitonic transitions, its relation to the magnetization, M, and to the effective g- factor, $\mathrm{g_{eff}}$, is given by \cite{gaj1993relation, bartholomew1986interband},

\begin{equation}
\begin{aligned}
   \Delta E_{z}^{sp-d} &= E_{+} - E{_{-}} &= \frac{\beta_{exc} - \alpha_{exc}}{g_{Mn} \, \mu_{B}} \, M = g_{eff} \, \mu_{B} \, B
\end{aligned}
\label{eqn:9}
\end{equation}

\begin{figure*}[htbp!]
    \centering
       \includegraphics[width=1.01\linewidth]{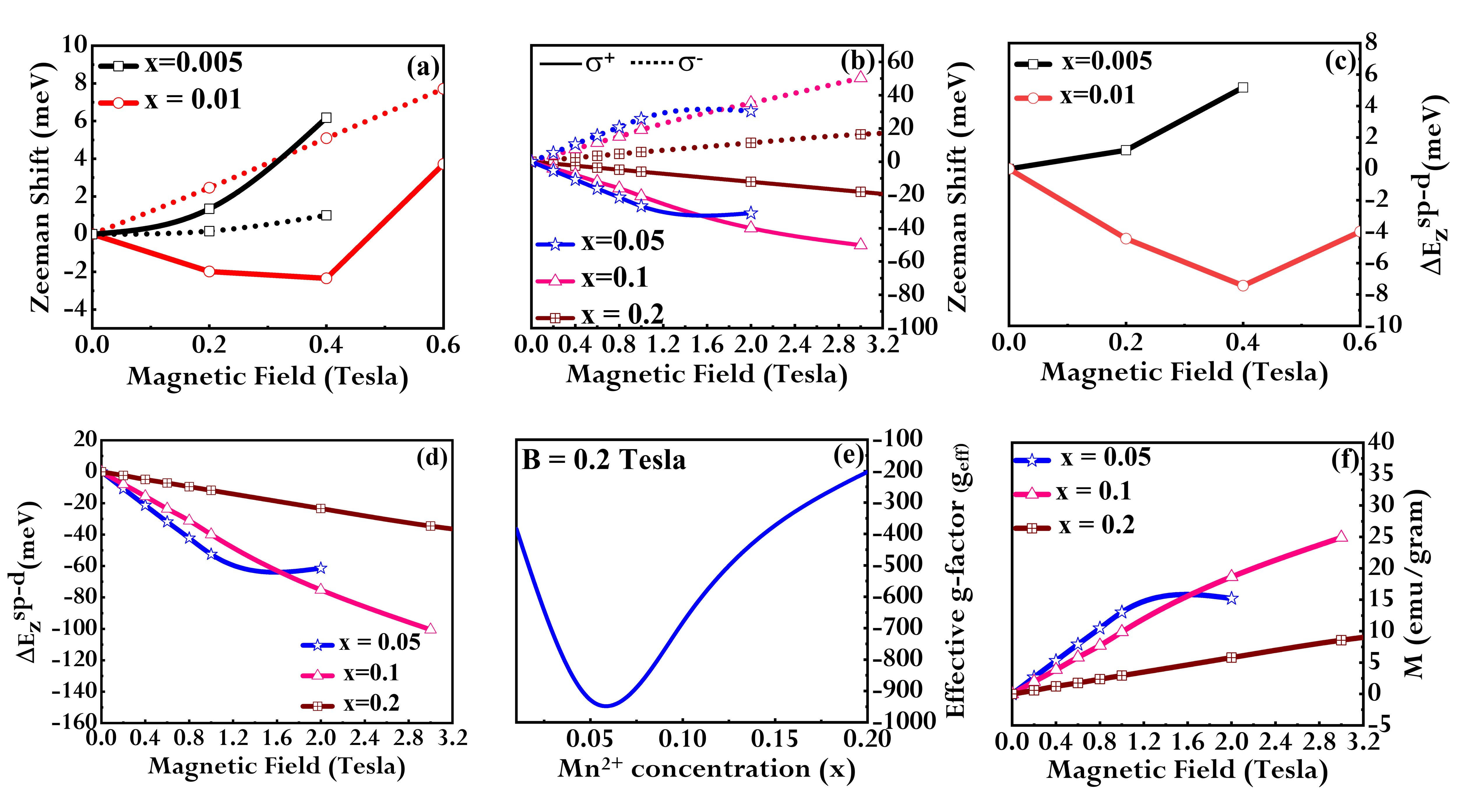}
    \caption{Zeeman shift related to zero field magneto-exciton energy, and Zeeman splitting energy ($\mathrm{\Delta E_{z}^{sp-d}}$) \textit{vs} magnetic field for various dopant concentrations. (a), (c) $\mathrm{x \leq 0.01}$, (b), (d) x > 0.01. (e) Concentration dependent effective g-factor ($\mathrm{g_{eff}} $) for a fixed strength of magnetic field, B = 0.2Tesla at T = 4.2K. (f) Magnetization (M) calculated using modified Brillouin function for $\mathrm{x \geq 0.05}$.}
       \label{Fig:3}
   \end{figure*}
   
\section{Results and Discussion \label{3}}
\subsection{Magnetic-field induced excitonic interband transition energy \label{3.1}}
\vspace{0.5cm}
\indent Magnetic field dependence of the PL transition energy ($\mathrm{E_{Tex}}$) for both $\mathrm{\sigma^{+}}$ and $\mathrm{\sigma^{-}}$ polarization is computed for ring dimensions R = 80{\AA}, d = 20{\AA}, which is approximately equals to the effective Bohr radius of the exciton, and the results are displayed for low ($\mathrm{x \leq 0.01}$) and high concentrations ($\mathrm{x > 0.01}$) of $\mathrm{Mn^{2+}}$ ions in Fig. \ref{Fig:2}. The transition energy increases with the increasing concentration of $\mathrm{Mn^{2+}}$ ion because the bandgap ($\mathrm{E_{g}}$) is directly proportional to the latter. The calculation using the above theoretical model shows that at B = 0, the PL is unpolarized, i.e., the energies of $\mathrm{\sigma^{\pm}}$ magneto-exciton are degenerate. However, the applied magnetic field breaks the degeneracy and causes the PL to split into left ($\mathrm{\sigma^{-}}$) and right ($\mathrm{\sigma^{+}}$) circularly polarized. This is indicated by a monotonic shift of $\mathrm{E_{Tex}}$ towards low and high energies about zero field energy in Fig. \ref{Fig:2}, and the PL gets resolved into two branches of exciton doublet corresponding to $\mathrm{\sigma^{+}}$ and $\mathrm{\sigma^{-}}$ polarization, respectively. The reason for this is attributed to the fact that the applied magnetic field influences the potential barrier height of the two different spin components in a unique way owing to the sp-d exchange interaction, as discussed in sec. \ref{2}.\\
\indent The variation of $\mathrm{E_{Tex}}$ with B for the QR doped with low $\mathrm{Mn^{2+}}$ concentration ($\mathrm{x \leq 0.01}$ in Fig. \ref{Fig:2}(a)) is different from the QR doped with high concentration (Fig. \ref{Fig:2}(b)). Instead of showing a rapid fall with the magnetic field as seen for higher concentration, the $\mathrm{\sigma^{+}}$ transition energy mimics the $\mathrm{\sigma^{-}}$ transition, indicating a change of the PL emission from right circular to left circular polarization. A vivid picture of this unusual behaviour for low 'x’ has been well explained in a DMS QD by Kai Chang et al \cite{chang2003magnetic}, which is ascribed to the tuning of the effective g-factor to zero with the increasing field when the order of Zeeman splitting due to sp-d exchange interaction is comparable to the order of intrinsic Zeeman splitting. The sign of the former is opposite to the latter. The presence of crossing between $\mathrm{\sigma^{+}}$ and $\mathrm{\sigma^{-}}$ transition energy in Ref \cite{chang2003magnetic} is missing here for x = 0.01 because the data has been plotted for the extended range of magnetic fields, including the type-II region in Ref \cite{chang2003magnetic}, whereas it is limited to the type-I region in the present work. Typically, the order of intrinsic Zeeman splitting is much smaller than the energy level splitting induced by the sp-d exchange interaction; hence, the former is neglected in the present calculation. The inset in Fig. \ref{Fig:2}(b) shows the high field data for x = 0.2.

\begin{figure*}[htbp!]
\begin {subfigure}
    \centering
    \includegraphics[width=17cm, height=10cm]{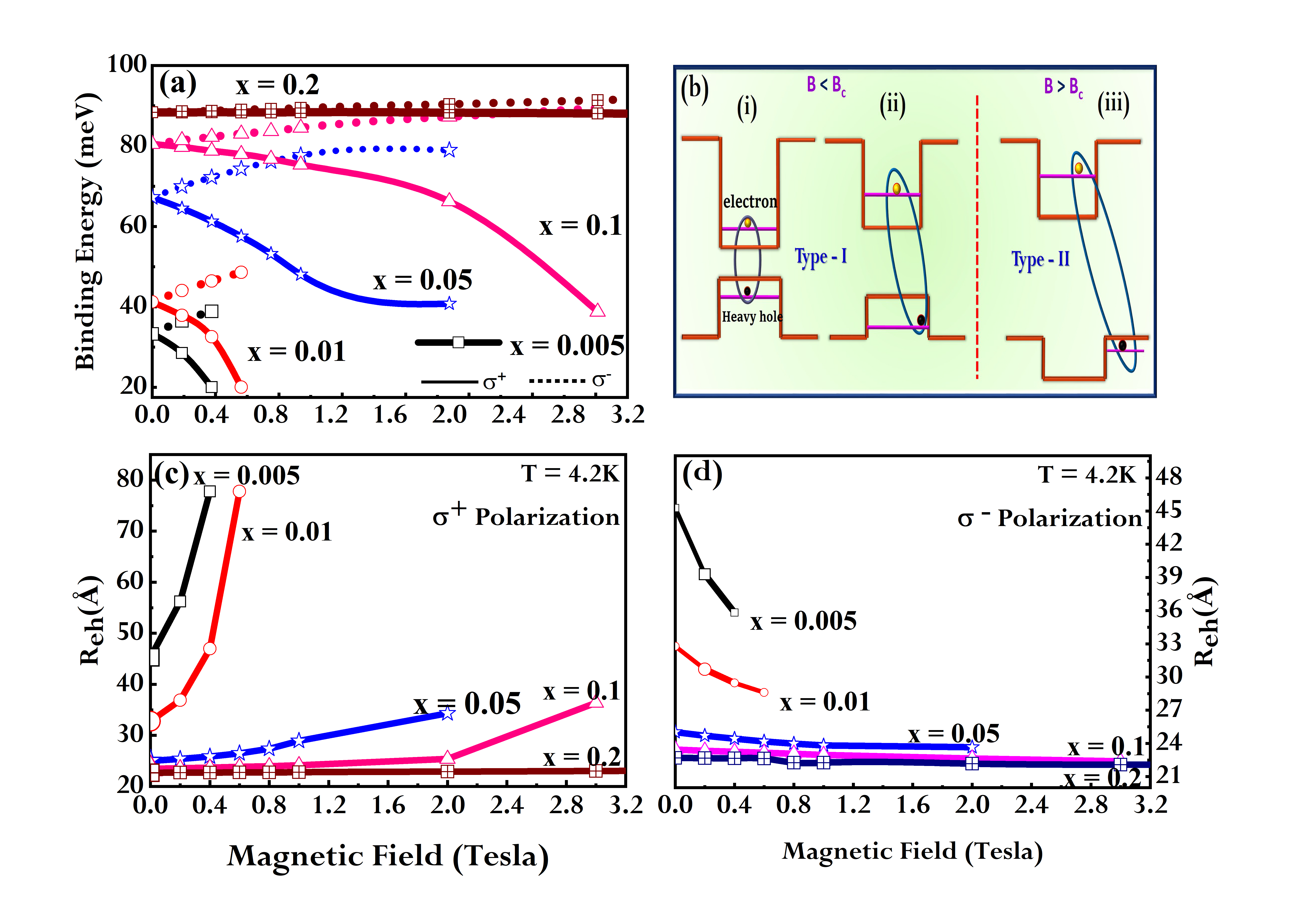}
   \caption{(a) Binding energy of $\mathrm{\sigma^{\pm}}$ magneto-exciton \textit{vs} magnetic field. (b) Schematic explaining the overlap integral between the electron and hole for various strengths of magnetic field: $\mathrm{(i) \, B = 0, (ii) \, 0 < B < B_{c}, \, and \, (iii) \, B > B_{c}}$. In-plane electron–hole distance corresponding to (c) $\mathrm{\sigma^{+}}$, and (d) $\mathrm{\sigma^{-}}$ exciton \textit{vs} magnetic field for various dopant concentrations.}
    \label{Fig:4}
    \end{subfigure}
\begin {subfigure}
     \centering
    \includegraphics[width=\linewidth]{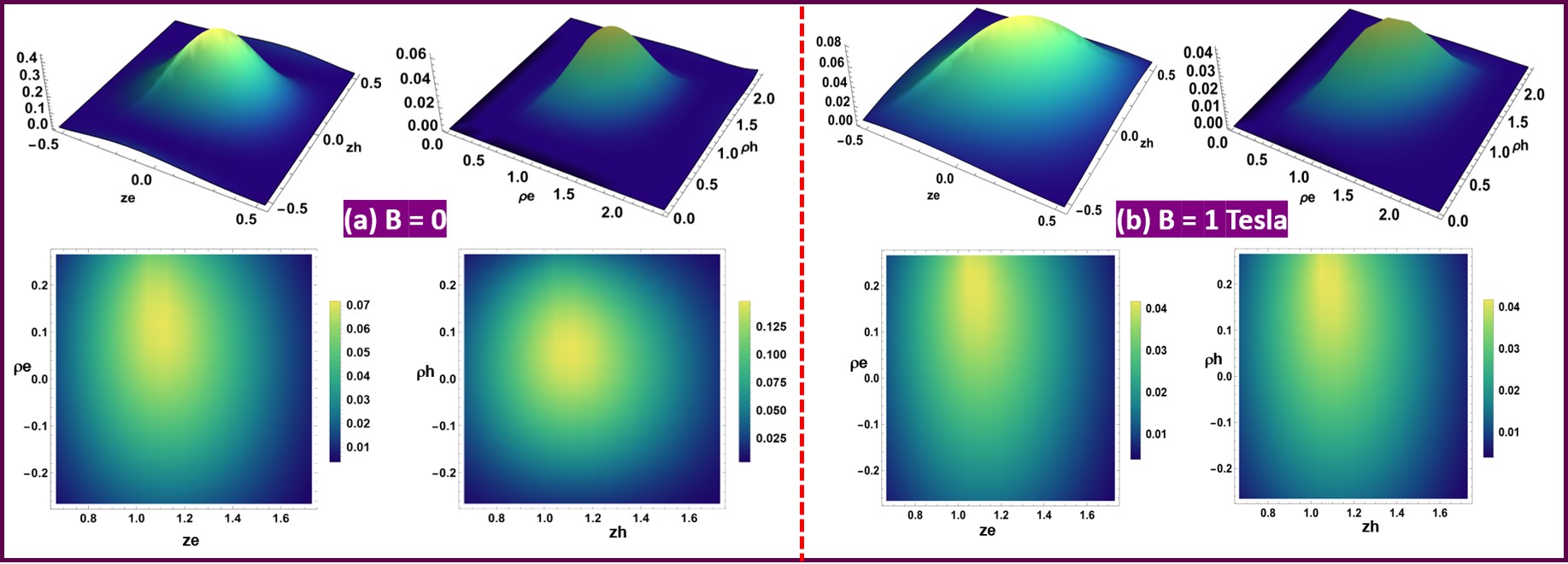}
    \caption{\textit{Upper panel}: 3D-plot for the probability distribution of electrons and holes along axial, and radial direction. \textit{Lower panel}: Density plot of the probability distribution of single-particle states along both radial and axial direction. The data has been plotted for (a) B = 0, and (b) B = 1Tesla.}
    \label{Fig:5}
\end{subfigure} 
\end{figure*}

\subsection{Zeeman shift and Zeeman splitting of the exciton energy levels \label{3.2}}
\vspace{0.5cm}
\indent Figures \ref{Fig:3}(a) and \ref{Fig:3}(b) plot the magnetic field dependence of the exciton transition energy as Zeeman shifts ($\mathrm{E_{ex} (B) - E_{ex} (B = 0))}$ relative to the zero-field exciton energy for both the transitions, which is also described by Eq. (\ref{eqn:8}). It is noted from figure that the shift increases with increasing magnetic field for both the transition, but it shows a positive and negative increment for $\mathrm{\sigma^{-}}$ and $\mathrm{\sigma^{+}}$ which corresponds to the blue and redshift in the interband transition energy (Fig. \ref{Fig:2}), respectively. Interestingly, one could observe the symmetric Zeeman splitting about the zero-field energy for the QR doped with 5\% and 10\% of $\mathrm{Mn^{2+}}$ ions, but for all other dopant concentrations (0.5\%, 1\%, and 20\%), the splitting seems to be asymmetric. \\
\indent On the quantitative footing, the Zeeman splitting energy, $\mathrm{\Delta E_{z}^{sp-d}}$, plotted in Fig. \ref{Fig:3}(c) and \ref{Fig:3}(d) is described as the energy difference between the two excitonic transitions under $\mathrm{B}$ and is determined from the data plotted in Fig. \ref{Fig:3}(a) and \ref{Fig:3}(b) as given in Eq. (\ref{eqn:9}). The magnetic field suppresses the $\mathrm{Mn^{2+}}$ spin fluctuations by aligning the randomly oriented $\mathrm{Mn^{2+}}$ spins along the field direction, indicating a state of magnetic ordering, thereby increasing $\mathrm{\left \langle S_{z}\right \rangle}$ and causing the GZS. It is interesting to note from Fig. \ref{Fig:3}(c) and \ref{Fig:3}(d) that $\mathrm{\Delta E_{z}^{sp-d}}$ increases with the dopant concentration up to x = 0.05, and thereafter it starts decreasing. This is because the Zeeman splitting is proportional to the effective dopant concentration ‘$\mathrm{x_{eff}}$’ as given in Eq. (\ref{eqn:8}), and the latter increases with ‘x’ and shows a maximum at a particular concentration. Henceforth, it starts to move downhill because of the antiferromagnetic interactions between the nearest neighbouring magnetic ions, which cancels the spins of the corresponding pairs and reduces the effective contribution to the thermal average of the spin polarization of $\mathrm{Mn^{2+}}$ ions, $\mathrm{\left \langle S_{z}\right \rangle}$. The strength of the Zeeman splitting can be directly evidenced from the absolute value of effective g-factor which has been calculated for various ‘x’, and is plotted in Fig. \ref{Fig:3}(e).\\
\indent Figure \ref{Fig:3}(f) shows the magnetization (M) \textit{vs} magnetic field curves for $\mathrm{x \geq 0.05}$. Magnetization increases with the magnetic field since it enhances $\mathrm{\left \langle S_{z}\right \rangle}$, showing a linear dependence on magnetic field, which is an expected paramagnetic behaviour in any CdMnTe based quantum systems. As already discussed, when QR is populated with more magnetic ions, the spin-spin interaction becomes more robust, which results in a quenching of magnetization for high ‘x’ because of the lower value of $\mathrm{\left \langle S_{z}\right \rangle}$.

\begin{figure*}[htbp!]
 \centering
     \includegraphics[width=17cm, height=10cm]{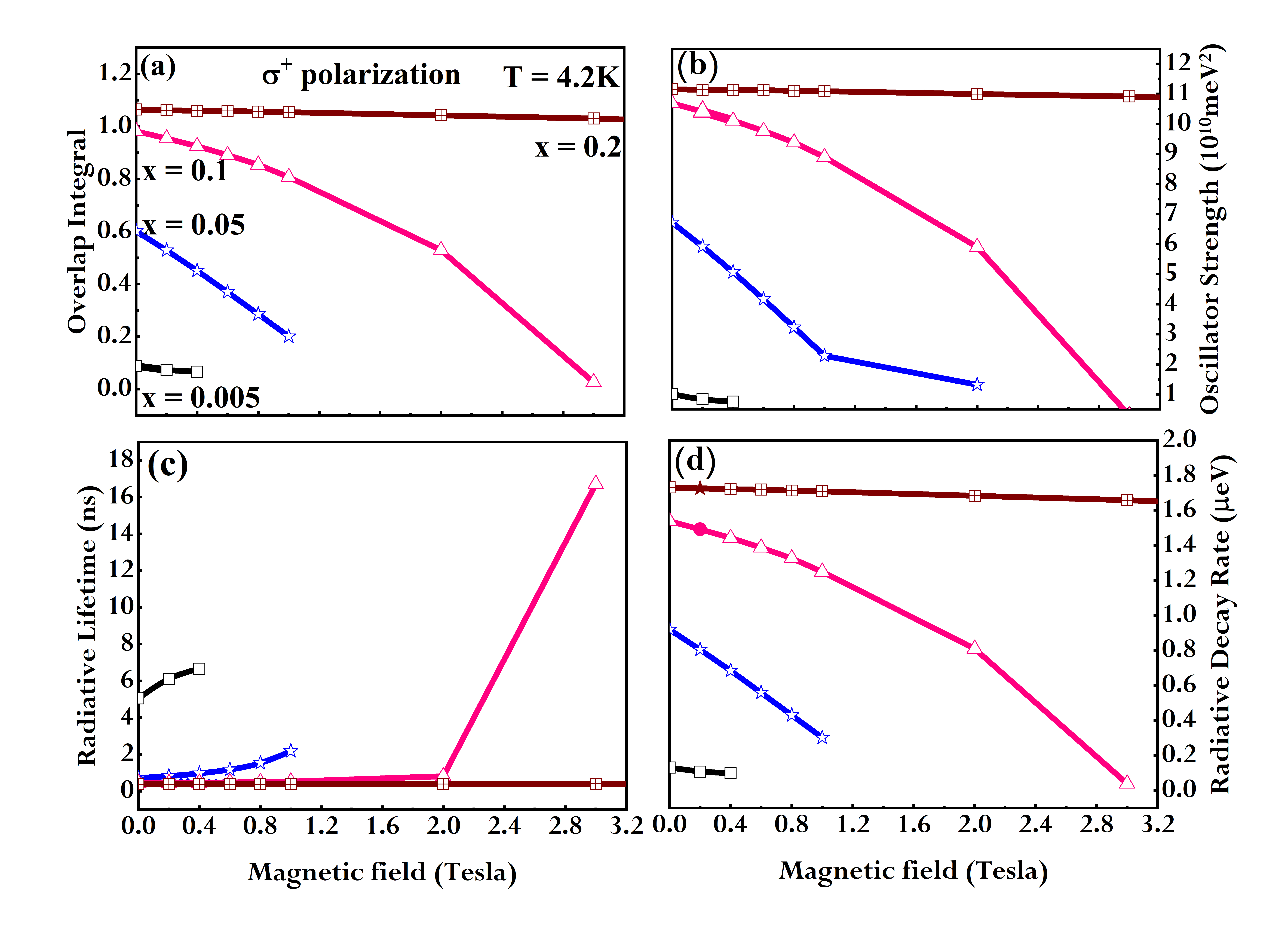}
    \caption{Variation of (a) overlap integral, (b) oscillator strength, (c) radiative lifetime, and (d) radiative decay rate with the magnetic field for $\mathrm{\sigma^{+}}$ and $\mathrm{\sigma^{-}}$ transitions.}
    \label{Fig:6}
   \end{figure*}

\subsection{Binding energy of $\mathrm{\sigma^{\pm}}$ magneto-exciton \label{3.3}}
\vspace{0.5cm}
\indent The variation of binding energy is plotted in Fig. \ref{Fig:4}(a) for various concentrations of $\mathrm{Mn^{2+}}$ ions. The trend of the binding energy for both $\mathrm{\sigma^{-}}$ and $\mathrm{\sigma^{+}}$ polarization concerning the magnetic field is as same as the trend followed by the interband transition energy, and this behaviour persists for different concentrations also. Nevertheless, for $\mathrm{\sigma^{+}}$ polarization, there is a rapid decrease of binding energy with the magnetic field as compared to the steady increase for $\mathrm{\sigma^{-}}$ polarization. This can be better understood from the schematic in Fig. \ref{Fig:4}(b), which explains how the applied magnetic field modifies the electron-hole overlap inside a SQR. \\
\indent At B = 0, the location of both the electron and hole is in the same CdTe layer (Fig. \ref{Fig:4}(b)(i)). Zeeman splitting of the energy levels in the valence band is highly sensitive to the applied field, which is not the case with the conduction band. This is because the band offset formed in the conduction band is generally larger than the valence band offset since 80\% of the bandgap difference falls in the former.  Moreover, the absolute value of the exchange constant, which represents the strength of the exchange interaction, is larger for the heavy hole ($\mathrm{|\beta_{exc} N_{0} = 880meV|}$) than for the electron ($\mathrm{|\alpha_{exc} N_{0} = 220meV|}$). Therefore the electron with $\mathrm{s_{z} = -1/2}$ in the conduction band would forever be confined in the non-magnetic CdTe layer itself irrespective of the strength of the applied field as its potential band offset is sufficiently larger than the order of magnetic splitting (Fig.~\ref{Fig:4}(b)(ii)). However, the potential barrier for the heavy hole with $\mathrm{J_{z} = -3/2}$ is tremendously reduced with the magnetic field, and it encounters a flat band situation at critical field value, beyond which the system undergoes a type-I - type-II transition (Fig. \ref{Fig:4}(b)(iii)). As a result, the electron remains in the CdTe layer, but the hole moves towards the heterostructure interface and finally to the CdMnTe layer. Hence, the exciton will no longer be spatially direct; rather, it becomes spatially indirect, which reduces the overlap between the electron and hole, whereby spin-down exciton states have reduced binding energy.
\indent To justify this discussion, the in-plane exciton radius, $\mathrm{R_{eh}}$, the average distance between the electron and hole in the plane of the QR, has been calculated and is plotted in Fig. \ref{Fig:4}(c) and \ref{Fig:4}(d). As anticipated, the monotonic increase and decrease of $\mathrm{R_{eh}}$ could be seen for $\mathrm{\sigma^{+}}$ and $\mathrm{\sigma^{-}}$ polarization, respectively, for all x. Moreover, the 3D plot of the probability distribution of spin-down electrons and holes ($\mathrm{|\Psi|^{2}}$) along $\mathrm{\rho}$ and z-directions of the QR, and the density plot of the single-particle distribution depicted in Fig.~\ref{Fig:5} helps to understand the effect of magnetic field on the carrier confinement inside the QR. Obviously, $\mathrm{|\Psi|^{2}}$ is larger for zero magnetic field as one can compare the order of magnitude between B = 0  and B = 1Tesla.

\subsection{Oscillator strength, radiative linewidth and radiative lifetime of magneto-exciton \label{3.4}}
\vspace{0.5cm}
To gain further insight into the $\mathrm{\sigma^{+}}$ and $\mathrm{\sigma^{-}}$ transition and related radiative properties, the investigation of oscillator strength (OS), radiative decay rate (RDR), and radiative lifetime (RLT) have been performed, and the results are delineated. The expression for the exciton oscillator strength follows \cite{nakamura1990faraday, ivchenko1992exciton, wu2012exciton},

\begin{equation}
\begin{aligned}
  &f_{\pm} =  \frac{E_{P}}{2 \, E_{T \pm}} I |\Omega (0)|^{2}\\
  &I = \Bigg |\int_{-\infty}^{+\infty} N_{1s} \, \phi_{e} \, (\rho_{e}) \, \phi_{h} (\rho_{e}) \, f_{e}( z_{e}) \, f_{h}(z_{e}) \, d\rho_{e} \, dz_{e} \Bigg |^{2}
\label{eqn:10}
\end{aligned}
\end{equation}

\noindent where, the Kane energy, $\mathrm{E_{P} = 2.1eV}$ for CdTe, and, $\mathrm{E_{T \pm}}$ represents the interband transition energy corresponding to $\mathrm{\sigma^{+}}$ and $\mathrm{\sigma^{-}}$ transitions, respectively. The OS mainly depends on the overlap integral ‘I’ between the electron and hole envelope wavefunctions, and $\mathrm{\Omega(0)}$ denotes the probability of finding the electron and hole at the same position. The oscillator strength per unit area is proportional to the effective Bohr radius as, $\mathrm{F_{\pm} = \frac{1}{a_{B}^*{}^2} f_{\pm}}$. Exciton radiative lifetime, ‘$\mathrm{\tau}$’ (radiative decay rate, ‘$\mathrm{\Gamma = 1/\tau}$’) can be related to OS according to \cite{fonoberov2003excitonic, sivalertporn2012direct},

\begin{equation}
\begin{aligned}
\tau &= \frac{2 \pi \epsilon_{0} m_{0} c^{3} \hbar^{2}}{n e^{2} E_{T \pm} f_{\pm}}
\end{aligned}
\label{eqn:11}
\end{equation}

\noindent Here, the fundamental physical constants have their usual meaning and ‘n’ represents the refractive index of the material CdTe. The evolution of the oscillator strength as a function of magnetic field solely depends on the spatial overlap between the electron and hole wave functions, which has been depicted in Fig. \ref{Fig:6}(a) and \ref{Fig:6}(b). The applied magnetic field increases the overlap between the electron and hole ground states for $\mathrm{\sigma^{-}}$ polarization, indicating larger OS due to the increase of potential barrier height. As expected for the $\mathrm{\sigma^{+}}$ polarization, the OS sensitively depends on B, which diminishes the excitonic effect by spatially separating the electron and hole as explained in sec. \ref{3.3} and thereby weakens the corresponding optical transition. The overlap integral increases as the dopant concentration increases due to the increased potential barrier height. \\
\indent Figure \ref{Fig:6}(c) and \ref{Fig:6}(d) shows the radiative lifetime and radiative decay rate as a function of magnetic field for various dopant concentrations. The RLT of exciton increases with increasing B for $\mathrm{\sigma^{+}}$ polarization, which is accompanied by a decrease in RDR. The exciton lifetime is found to decrease from 5.04ns to 0.38ns when the concentration of $\mathrm{Mn^{2+}}$ ion increases from x = 0.005 to x = 0.2 at B = 0, where radiative recombination dominates (Fig. \ref{Fig:6}(c)). The RDR, which characterizes the decay of photon emitted by the exciton, shows its maximum only for B = 0, which elucidates the probability of finding an electron and hole at the same position ($\mathrm{r_{e} = r_{h}}$) is more prominent in the absence of magnetic field.

\begin{figure*}[htbp!]
 \centering
       \includegraphics[width=\linewidth, height=6cm]{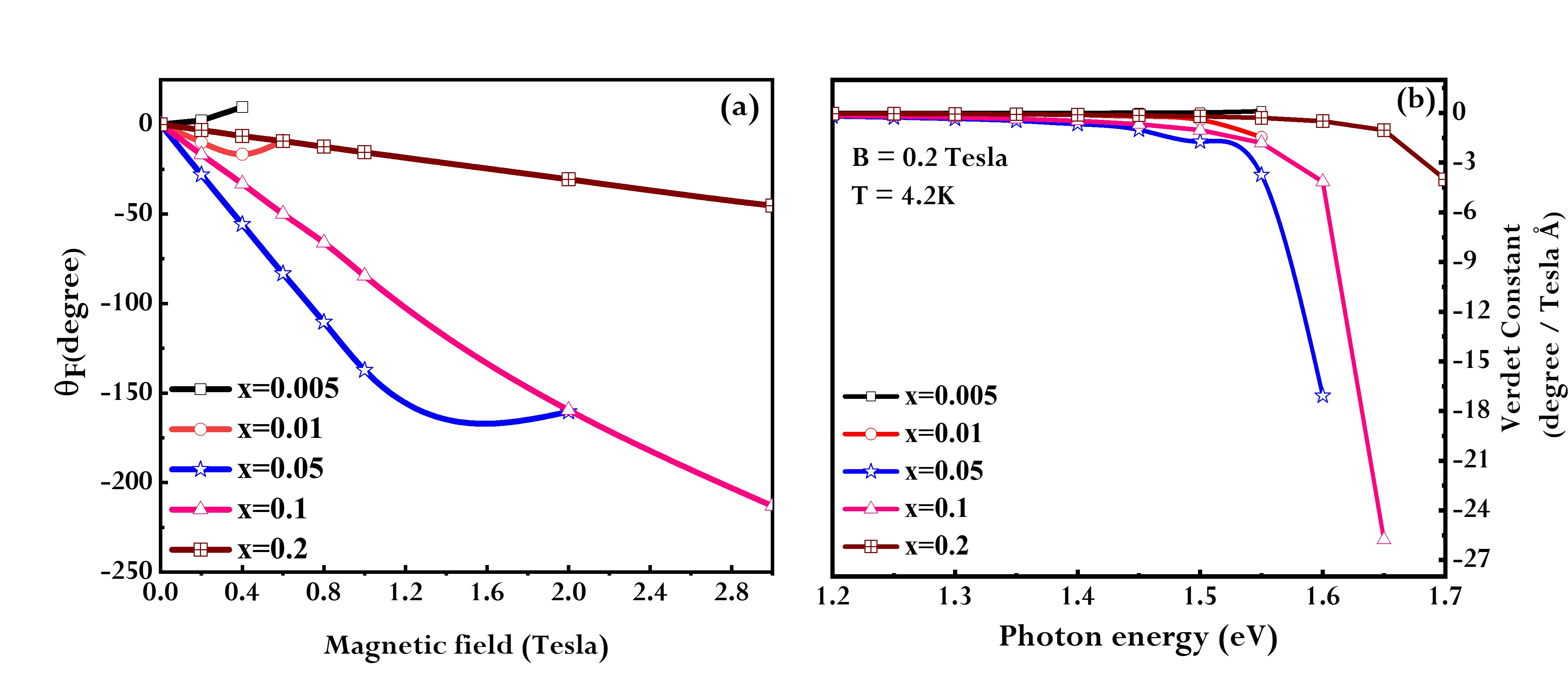}
 \caption{(a) Faraday rotation angle ($\mathrm{\Theta_{F}}$) as a function of magnetic field for a fixed photon energy of $\mathrm{E_{ph}=1.5eV}$, and (b) Verdet constant as a function of photon energy for various dopant concentrations for a fixed strength of magnetic field, B = 0.2Tesla.}
    \label{Fig:7}
   \end{figure*}
   
\subsection{GFR in semimagnetic SQR \label{3.5}}
\vspace{0.5cm}
\indent The Faraday rotation (Fig. \ref{Fig:1}(c)), results from a difference in refractive indices of the left and right circularly polarized light after traveling through a magnetized medium with a length ‘$\mathrm{\iota}$’. The phase difference in velocity between the two circularly polarized components is expressed through the FR angle as \cite{bartholomew1986interband},

\begin{equation}
\begin{aligned}
 \Theta_{F} &= \frac{\Delta\phi}{2} &= \frac{E l}{2 \hbar c} \left(n_{-} - n_{+}\right)
 \end{aligned}
\label{eqn:12}
\end{equation}

\noindent Here, $\mathrm{n_{-}}$ and $\mathrm{n_{+}}$ denote the refractive indices of the left and right circular polarized light, and E is the incident photon's energy. As aforementioned, the FR in DMS alloys is a giant one due to the large Zeeman splitting of the energy levels as a result of sp-d exchange interaction, which has been computed using the single oscillator model as preferred in the work of Bartholomew et al., After performing a series of calculations, $\mathrm{\Theta_{F}}$ achieves the final form as \cite{bartholomew1986interband},
\begin{equation}
\begin{aligned}
\Theta_{F} &= \frac{\sqrt{F_{0}} l}{2 \hbar c} \left(\frac{\beta_{exc} - \alpha_{exc}}{g_{Mn} \, \mu_{B}}\right) \, M \,  \frac{1}{E_{0}} \frac{y^{2}}{(1-y^{2})^{3/2}} \, ; \, y = \frac{E}{E_{0}}
\end{aligned}
\label{eqn:13}
\end{equation}

\noindent Here, $\mathrm{F_{0}}$ is a constant that involves the oscillator strength, and $\mathrm{E_{0}}$ is the ground state interband transition energy at the fundamental energy gap at zero magnetic field. The angle is directly proportional to the GZS through the term $\mathrm{\Delta E = \frac{\beta_{exc} - \alpha_{exc}}{g_{Mn} \mu_{B}} M}$. The Verdet constant is written as the Faraday rotation per unit magnetic field per unit length, which is defined as \cite{bartholomew1986interband},

\begin{equation}
\begin{aligned}
V_{d} (E)&=\frac{\Theta_{F}}{B l} &= \frac{\sqrt{F_{0}}}{2 \hbar c} \, \left(\frac{\beta_{exc} - \alpha_{exc}}{g_{Mn \,} \mu_{B}}\right) \, \frac{\partial M}{\partial H} \,  \frac{1}{E_{0}} \, \frac{y^{2}}{(1-y^{2})^{3/2}}
\end{aligned}
\label{eqn:14}
\end{equation}

\indent Figure \ref{Fig:7}(a) depicts $\mathrm{\Theta_{F}}$ for the DMS QR doped with dilute, arbitrary, and high 'x' at a fixed photon energy of 1.5eV. It is noted from figure that the rotation angle increases with the increasing magnetic field since the applied field enhances the Zeeman splitting. The variation of the Verdet constant with the incident photon energies for a fixed magnetic field of B = 0.2Tesla is plotted in Fig. \ref{Fig:7}(b). The Verdet constant shows a sharp increase whenever the band gap resonance occurs (when the energy of the incident photon approaches the absorption edge of the material), and the photon energy, at which the Verdet constant shows a rapid enhancement, shifted to higher energies for the heavily doped QR because the absorption edge increases as the concentration of $\mathrm{Mn^{2+}}$ ions increases. Though the single oscillator model yields gratifying results, in which the behaviour of $\mathrm{E_{0}}$ at $\mathrm{\Gamma}$ point has been crudely modelled as constant at all temperatures, the success of using this in QR could not be verified due to a lack of reliable experimental data.
\section{Concluding Remarks \label{4}}
\vspace{0.5cm}
\indent Probing the exciton energy states in an applied magnetic field has been studied in semimagnetic QR, and the theoretical investigation of tuning related MO properties has been attempted. It is found that the doubly-connected topological structures like QR provide robust confinement for the carriers compared to single-connected topological QDs \cite{gnanasekar2004spin}. The difference in the behaviour of magneto-exciton energies between the QR doped with low and high $\mathrm{Mn^{2+}}$ ion concentrations has been explained in detail. The results show pronounced excitonic Zeeman splitting for low ‘x’ than high ‘x’, where the possibilities for the manganese ions to form antiferromagnetic pairs in the latter case are maximized. Among all the concentrations discussed here, x = 0.05 exhibits larger Zeeman splitting with the absolute value of effective g-factor, $\mathrm{g_{eff} = 928}$ (Fig. \ref{Fig:3}(e)), which gives rise to ultra-high Verdet constant of -15 degree/Tesla/\AA \; ($\mathrm{2.6 \times 10^{9} rad/Tesla/m}$), and the latter is $\mathrm{10^{4} - 10^{6}}$ orders of magnitude larger than in bulk $\mathrm{Cd_{1-x}Mn{x}Te}$ \cite{bartholomew1986interband, gaj1993giant, hugonnard1994faraday}, thin films \cite{masterson1997investigation, koyanagi1987epitaxial, shuvaev2011giant}, and is $\mathrm{10^{2}}$ orders larger than in QWs \cite{nakamura1990faraday, buss1995excitonic}, superlattices \cite{kohl1991faraday, nakamura1992faraday} as reported in the previous studies. This elucidates the importance of DMS-based QR in MO devices operating at a wavelength shorter than $\mathrm{1 \mu m}$ than already existing MO materials, such as Yttrium Iron Garnet (YIG) and Terbium Aluminum Garnet (TAG), organic molecules, conjugated polymers \cite{bartholomew1986interband, carothers2022high, kumari2018study}.\\ 
\indent Moreover, the low-temperature exciton lifetime is 715ps, whereas it is $\mathrm{\approx 100ps}$ in QWs doped with 25\% $\mathrm{Mn^{2+}}$ ion concentration \cite{polhmann1992exciton}. The study of exciton lifetime in semimagnetic quantum systems is impressive since it affects the optical properties and the magnetization dynamics of the concerned systems to a greater extent. The exciton lifetime in DMS determines the formation of bound magnetic polaron (BMP) \cite{harris1983formation, kalpana2017bound} or exciton magnetic polaron (EMP) \cite{akimov2017dynamics}, which causes spontaneous ferromagnetic ordering even in the absence of an external magnetic field due to the strong sp-d exchange interaction. Since the recombination limits the exciton lifetime, it interrupts the EMP formation before the polaron reaches its stable state. If the exciton does not decay during the process of EMP formation, then the EMP would reach its equilibrium state, which is accompanied by a decrease of exciton energy and provides an additional localization for the carriers. The reliability of the results obtained using single oscillator model could not be verified due to the missing experimental data, but it is believed to be improved using the multi oscillator model as adopted in \cite{jimenez1992near}. Since the low path length and the modest magnetic field yields a ultra-high Verdet constant, theoretical demonstration of generating larger FR and higher Verdet constant in DMS QRs would incite interest in preparing high-quality QR heterostructures based on DMS.\\
\indent With the unrivaled ability to modulate the magnetic excitonic transitions and thereby the optical activity of the materials at the nanoscale for a broader energy spectrum with various mole fractions of $\mathrm{Mn^{2+}}$ ions in external magnetic fields and effective magnetic switching of the spins make DMS-based QR a judicious choice among promising candidates for applications in future spin-photonic and spin-electronic devices.

\bibliography{main}

\end{document}